\documentclass[11pt]{article}

\usepackage[french,english]{babel}
\usepackage[T1]{fontenc}

\usepackage{amsfonts,amsthm}
\usepackage{amsmath,mathtools}   
\usepackage{latexsym}
\usepackage{amssymb}
\usepackage{mathrsfs}
\usepackage{pifont}
\usepackage{tikz,ulem}
\usepackage[all]{xy}
\usepackage{hyperref}

\newtheoremstyle{newrem}{3pt}{3pt}{}{}
{\bfseries}{.}{.5em}{}

\newtheorem{theo}{Theorem}[section]

\newtheorem*{theo*}{Theorem}

\newtheorem{prop}[theo]{Proposition}

\newtheorem{coro}[theo]{Corollary}

\theoremstyle{newrem}

\theoremstyle{definition}

\newtheorem*{term*}{Notation/Terminology}

\def\sR{{\mathscr{R}}}

\def\osR{{\overline{\mathscr{R}}}}

\def\ocV{\overline{{\mathcal{V}}}}

\def\ocX{\overline{{\mathcal{X}}}}

\def\ocZ{\overline{{\mathcal{Z}}}}

\def\cV{{\mathcal{V}}}

\def\cX{{\mathcal{X}}}

\def\cZ{{\mathcal{Z}}}

\def\fc{{\mathfrak{c}}}

\def\ff{{\mathfrak{f}}}

\def\fR{{\mathfrak{R}}}

\def \oA{{\overline{A}}}
\def \oB{{\overline{B}}}

\def \oW{{\overline{W}}}
\def \oX{{\overline{X}}}
\def \oZ{{\overline{Z}}}

\def\ba{{\mathbf{a}}}
\def\bb{{\mathbf{b}}}

\def\bd{{\mathbf{d}}}
\def\be{{\mathbf{e}}}

\def\un{{\mathbb{I}}}

\usepackage{geometry}
\title{\bf Algebras behind the bispectrality of \\the Wilson rational functions and their ${}_4\phi_3$ limits}
\author{
Nicolas Cramp\'e (Corresponding author)\textsuperscript{$1$},
Satoshi Tsujimoto\textsuperscript{$2$},\\
Luc Vinet\textsuperscript{$3,4$},
Alexei Zhedanov\textsuperscript{$2$}
\\[.9em]
\textsuperscript{$1$}
\small CNRS -- Universit\'e de Montr\'eal CRM - CNRS,
\\
crampe1977@gmail.com\\[.5em]
\textsuperscript{$2$}
\small Graduate School of Informatics, Kyoto University
Yoshida-Honmachi, Kyoto, Japan 606-8501,\\[.5em]
 \textsuperscript{$3$}
\small Centre de Recherches Math\'ematiques, Universit\'e de Montr\'eal, P.O. Box 6128, \\
\small Centre-ville Station, Montr\'eal (Qu\'ebec), H3C 3J7, Canada.\\[.9em]
\textsuperscript{$4$}
\small IVADO, Montr\'eal (Qu\'ebec), H2S 3H1, Canada.\\[.9em]
\textsuperscript{$4$}
\small
School of Mathematics, Renmin University of China, Beijing 100872, China\\[.9em]
}
\date{}

\bigskip\bigskip

\begin{document}
\maketitle
\begin{abstract}
The properties of the Wilson rational functions ${}_{10}\phi_9$
  with three different normalizations are described. For one normalization, it satisfies an $R_{II}$ recurrence relation, whereas for the two other ones, they satisfy a generalized eigenvalue problem. The so-called Wilson rational algebra is introduced, which encodes algebraically the spectral properties of these special functions. Finally, different limits are considered, leading up to functions proportional to  ${}_{4}\phi_3$. For one of these, the spectral algebra simplifies to yield the meta $q$-Racah algebra.
\end{abstract}
\vspace{0.5cm}
\textbf{Keywords:} Hypergeometric functions; Recurrence relations; Bispectrality; Generalized eigenvalues problem.
\vspace{0.5cm}

\section{Introduction}

This paper explores the algebraic structures that underlie the bispectrality of families of biorthogonal rational functions (BRF) of hypergeometric type. The adjective bispectral applies to functions that depend on two sets of variables say, "coordinates" and "degrees" and obey two groups of spectral equations one involving operators acting on the coordinates with eigenvalues depending on the degrees and the other equations working in the reciprocal way. This line of research for BRF is in the spirit of the interpretation of the orthogonal polynomials (OP) of the Askey scheme in terms of the Askey-Wilson algebra \cite{zhedanov1991hidden}, \cite{crampe2021askey} and its specializations. In the case of the univariate OPs, bispectrality stems from the three-term recurrence relation and the differential or difference equation that the classical polynomials verify. Both equations are of the standard eigenvalue form. The underlying algebra is identified from the relations realized by the operators intervening in these equations and taken in either the coordinate or the degree representation. For finite families of polynomials, this is intimately related to the theory of Leonard pairs \cite{terwilliger2003introduction} which consists in two linear maps that each act on an eigenbasis for the other in a tridiagonal fashion. The corresponding bispectral OPs arise in the overlaps between the two bases of the operators forming the Leonard pair. These studies have much enriched the fruitful interface between orthogonal polynomials, special functions and representation theory and have generated advances in numerous other fields. The present article is part of the efforts aimed at determining how the features of bispectral biorthogonal rational functions can be algebraically encoded and unraveled.

Some of us have hence initiated explorations in those directions by first looking at the bispectral equations of (finite families of) rational functions of Hahn \cite{tsujimoto2021algebraic} and $q$-Hahn \cite{bussiere2022bispectrality} types. It is known from \cite{zhedanov1999biorthogonal} that biorthogonal rational functions appear as solutions of generalized eigenvalue problems (GEVP) involving two matrices acting tridiagonally. What emerged from the examination of the BRF of Hahn and $q$-Hahn types is that their bispectrality expressed itself through the fact that they verify two GEVPs set up from using in each case two operators from a triplet that has a role analogous to the Leonard pair for OPs. The algebra generated by the members of this triplet was identified and referred to as the rational Hahn or $q$-rational Hahn algebra. It was further observed \cite{vinet2021unified} that both the Hahn and the rational Hahn algebras admitted embeddings in a simpler algebra that was called the meta Hahn algebra which hence allowed to provided a unified and simultaneous characterization of both the Hahn OPs and BRFs. The same was found to hold for the $q$-analogs. These meta Hahn and $q$-meta Hahn algebras prove to have nice features: they have simple presentations in terms of generators and relations, they admit a two-generated subalgebra of Artin-Shelter type \cite{artin1987graded} and in addition to the embeddings mentioned above, their representations in the bases associated to the GEVP defined by their generators have remarkable properties. This leads one to believe that meta algebras can be associated to the various entries of the Askey scheme.

This prompted a systematic program which is underway \cite{tsujimoto2024meta}, \cite{bernard2024meta} and that aims to extend the Askey tableau by adjoining to the OP families the corresponding bispectral BRF and to provide the characterization of these functions from the representation theory of the associated meta algebras. As in the case of the orthogonal polynomials, this undertaking proceeds in two complementary ways. On the one hand, algebras are abstracted from the bispectral properties of the functions; on the second hand once the abstract algebras have been identified, one aims to show that the various representations series of these meta algebras (including the continuous ones) will elegantly capture and synthesize the essence of the special functions of interest. At this juncture, the identification of the meta Racah algebra and of its $q$-deformation has not been done and this is one of the goals pursued here by following the first route.

Soon after the discovery of the Askey-Wilson polynomials, a generalization of these polynomials to a biorthogonal rational function in terms of a balanced, very-well-poised ${}_{10}\phi_9$ has been introduced \cite{wilson1991orthogonal}, \cite{rahman1991biorthogonality}, \cite{rahman1993classical}, \cite{spiridonov2000spectral}. They are known as the Wilson functions and are bispectral satisfying a pair of GEVP. One of these GEVP can in fact be cast in the framework of the rational Heun operators of classical type \cite{tsujimoto2023rational}. The BRFs of $q$-Hahn type have been obtained in \cite{bussiere2022bispectrality} as a limit and specialization of the Wilson function and five families of BRFs of  ${}_{4}\phi_3$ form have previously been derived in this way in \cite{GM}. 

It is therefore appropriate to revisit the bispectral properties of the Wilson rational function to set the stage for our study of the algebras underneath the bispectrality of the BRFs of $q$-Racah type and this will be done in Section \ref{sect:2}. Attention will be paid to the fact that there is some freedom in the choice of the coefficients of GEVP and that the spectral equations depend on the normalization factor that is pulled out of the ${}_{10}\phi_9$ to define the function of interest. The properties of the standard Wilson rational function $W_n(x)$ \eqref{eq:qRacahRF} will be reviewed first and another rational version $\oW_n(x)$ \eqref{Wbar} will be introduced and shown to obey a second type of recurrence and difference GEVP.  The $R_{II}$ recurrence relation of the polynomials \eqref{RII} associated to $W_n(x)$ will also be recorded. The remainder of the paper will then proceed as follows. The rational Wilson algebra is identified in Section \ref{section:3} by determining the relations between the matrices $X$ and $Z$ that enter in the GEVP together with a diagonal matrix $V$. Remarkably the algebras arising in this fashion from $W_n(x)$ and $\oW_n(x)$ are structurally identical. Limits to ${}_4\phi_3$ are considered in Section \ref{section:4}. While all results presented in \cite{GM} are discussed, of particular interest are the limits of $W_n(x)$ and $\oW_n(x)$ that lead to rational functions of $q$-Racah type. In these two cases, the algebra that descends from the rational Wilson algebra has a remarkably simple structure and yields the meta $q$-Racah algebra whose identification was a chief purpose of this study. Interestingly, the limits stemming from $W_n(x)$ and $\oW_n(x)$ lead to the same meta algebra up to an exchange of the generators. Concluding remarks will be found in Section \ref{section:5}.

\section{Wilson rational functions} \label{sect:2}

The Wilson rational functions are defined in terms of the very-well-poised $q$-hypergeometric functions as follows, for $n$ a non-negative integer:
\begin{align}\label{eq:qRacahRF}
W_n(x)={}_{10}\phi_9 \left({{a,\;q\sqrt{a},\;-q\sqrt{a},\; q^{-n},\; bq^{n+1},\; q^{-x},\;  c  q^{x+1} ,\; ad  ,\;ae,\;af }\atop
{\sqrt{a},\;- \sqrt{a},\; aq^{-n}/b ,\;aq^{n+1},\; a q^{-x}/ c  ,\;   aq^{x+1},\;  q/d,\; q/e,\; q/f }}\;\Bigg\vert \; q;q\right)\,,
\end{align}
with the relation between the parameters
\begin{align}
 b   c  def  =1\,.
\end{align}
We suppose from now on that the previous constraint is satisfied.
The standard notations are used for the $q$-hypergeometric functions \cite{GR} defined by, 
\begin{align}
{}_{r+1}\phi_r\left({{a_1,\; a_2,\; \cdots,\; a_{r},\; q^{-n} }\atop
{b_1,\; b_2,\; \cdots,\; b_{r} }}\;\Bigg\vert \; q;z\right)=\sum_{k=0}^{n}
\frac{(a_1,a_2,\cdots,a_r,q^{-n};q)_k}{(b_1,b_2,\cdots,b_r,q;q)_k}z^k\,,
\end{align}
for $r,n$ non-negative integers ,
and where the $q$-Pochhammer symbols are
\begin{align}
(b_1,b_2,\cdots,b_r;q)_k=(b_1;q)_k(b_2;q)_k\cdots (b_r;q)_k\,,\qquad (b_i;q)_k=(1-b_i)(1-qb_i)\cdots (1-q^{k-1}b_i)\,.
\end{align}
For later convenience, let us introduce the coefficients
\begin{subequations} \label{eq:Z}
 \begin{align}
& Z_{n+1,n}{(b)}=\frac{(1-b q^{n+1}) (a-b q^{n+1})(d- q^{n+1} )(e- q^{n+1})(f- q^{n+1})}
 {(1-a q^{n+1}) (1-b q^{2n+2})(1-b q^{2n+1})}\,,\\
 &Z_{n-1,n}{(b )}=\frac{q(1-q^n)(1-aq^n)(1- d b  q^{n})(1- e b  q^{n})(1- f b  q^{n})}
 {(a-b q^n)(1-b q^{2n+1})(1-b q^{2n})}\,,
\end{align}
\end{subequations}
and
\begin{align}
&\lambda(q^x;c)=(1-q^{-x})(1- c  q^{x+1})\,,\\
&A_n(x)= Z_{n+1,n}{(b )}\underbrace{(1-aq^{x+n+1})(cq^{-n}/a-q^{-x} )}_{\displaystyle =\lambda(q^x;c) - 
\lambda(q^{-n-1}/a;c) } \,,\\
&B_n(x)= Z_{n-1,n}{(b )}\underbrace{(aq^x-bq^n)(q^{-n-x}/b-cq/a)}_{\displaystyle =\lambda(q^x;c) - 
\lambda(bq^n/a;c)} \,.
\end{align}
For $Z_{n+1,n}{(b )}$ and $Z_{n-1,n}{(b )}$, only the dependence on $b$ is specified in order to lighten the notations. 

By specializing theorem 2.4 of \cite{GM}, one can show that the Wilson rational functions \eqref{eq:qRacahRF} satisfy\footnote{The notations of the Section 4 of \cite{GM} has been modified as follows: $a\to a$, $b\to 1$, $s\to q^2 b$, $d\to a d$, $e\to ae$, $f\to af$, $e^{\xi}\to q^{x}$, $\mu \to q c $.}
\begin{align}\label{eq:GM}
 A_n(x) W_{n+1}(x)   -\Big(A_n(x)+B_n(x)+\lambda(q^x;c)\ba_n(b)\Big) W_{n}(x)+ B_n(x) W_{n-1}(x)=0\,,
\end{align}
where 
\begin{align}
 &\ba_n(b)= \frac{ q^{n+1} (1-b /a)(1-ad)(1-ae)(1-af)}{(1-a q^{n+1})(a-b q^n)}\,.
\end{align}

\paragraph{Generalized eigenvalue problem.}

As mentioned in \cite{Ros}, the Wilson rational functions satisfy a GEVP. We recall this property in the following proposition.
\begin{prop}
 The Wilson rational functions $W_n(x)$ satisfy the following GEVP, for $x,n$ non-negative integers: 
\begin{align}
 &X_{n+1,n}(b,c)\, W_{n+1}(x)  +X_{n,n}(b,c)\, W_{n}(x)+ X_{n-1,n}(b,c)\ W_{n-1}(x)\nonumber\\
=&\lambda(q^x;c)\Big( Z_{n+1,n}{(b )}\, W_{n+1}(x)+Z_{n,n}{(b)}\, W_{n}(x)    + Z_{n-1,n}{(b )}\, W_{n-1}(x)\Big)\,, \label{eq:GEVPW1}
\end{align}
where 
\begin{align}
&Z_{n,n}{(b )}=-(Z_{n+1,n}{(b )}+Z_{n-1,n}{(b)}+\ba_n(b))\,,
\end{align}
with $Z_{n,n+1}{(b )}$ and $Z_{n-1,n}{(b)}$ given by \eqref{eq:Z} and
\begin{align}
& X_{n+1,n}(b,c)=Z_{n+1,n}{(b )}\, \lambda(q^{-n-1}/a;c) 
\,,\\
& X_{n-1,n}(b,c)=Z_{n-1,n}{(b )}\,\lambda(bq^n/a;c)\,,\\
& X_{n,n}{(b,c )}= -\big(X_{n+1,n}(b,c)+X_{n-1,n}(b,c)\big)\,.
\end{align}
\end{prop}
\proof The proof is a rearranging of the terms in \eqref{eq:GM}.
\endproof

Let us emphasize that there is some freedom in the definitions of the coefficients $X$ and $Z$. Indeed, we can multiply \eqref{eq:GEVPW1} by any function depending on $n$ and redefine the terms such that $W_n(x)$ satisfy a GEVP with coefficients multiplied with this factor. We can also make the following affine transformations on the coefficients
\begin{align}\label{eq:Xtilde}
\widetilde X_{n+\epsilon,n}(b,c) =X_{n+\epsilon,n}(b,c) + \rho Z_{n+\epsilon,n}(b)\,,\quad \text{for } \epsilon \in \{0,1,-1\}\,.
\end{align}
In the case $\rho=-(1+qc)$, the GEVP takes the following form:
\begin{align}
 &\widetilde X_{n+1,n}(b,c)\, W_{n+1}(x)  +\widetilde X_{n,n}(b,c)\, W_{n}(x)+ \widetilde X_{n-1,n}(b,c)\ W_{n-1}(x)\nonumber\\
=& -(q^{-x}+cq^{x+1}) \Big( Z_{n+1,n}{(b )}\, W_{n+1}(x)+Z_{n,n}{(b)}\, W_{n}(x)    + Z_{n-1,n}{(b )}\, W_{n-1}(x)\Big)\,.\label{eq:GEVPWnorm1}
\end{align}

The previous proposition concerns a recurrence GEVP for the Wilson rational functions but a difference GEVP can be easily deduced.
\begin{coro}
 The Wilson rational functions $W_n(x)$ satisfy the following GEVP, for $x$ and $n$ non-negative integers: 
\begin{align}
 &X_{x+1,x}(c,b)\, W_{n}(x+1)  +X_{x,x}(c,b)W_{n}(x)+ X_{x-1,x}(c,b)W_{n}(x-1)\nonumber\\
=&\lambda(q^n;b)\Big( Z_{x+1,x}{( c )}\, W_{n}(x+1)+Z_{x,x}(c)\, W_{n}(x)    + Z_{x-1,x}(c)\, W_{n}(x-1)\Big)\,.\label{eq:GEVPW2}
\end{align}
\end{coro}
\proof Remark that the Wilson rational functions satisfy the following symmetry relation:
\begin{align}
 W_{x}(n)=W_{n}(x)\Big|_{b \leftrightarrow  c }\,.
\end{align}
Relation  \eqref{eq:GEVPW2} is proven applying this symmetry to \eqref{eq:GEVPW1}.\endproof

Let us remark that there exists a relation between the RHS of the recurrence and difference GEVP given by, for $n,x$ non-negative integers,
 \begin{align}\label{eq:ZZs}
&  Z_{n+1,n}(b)\, W_{n+1}(x)+Z_{n,n}(b)\, W_{n}(x)    + Z_{n-1,n}(b)\, W_{n-1}(x)\\
=&
 Z_{x+1,x}(c)\, W_{n}(x+1)+Z_{x,x}(c)\, W_{n}(x)    + Z_{x-1,x}(c)\, W_{n}(x-1)\;.\nonumber
\end{align}
for $n$ and $x$ non-negative integers. This relation has been proven by using explicit expressions \eqref{eq:qRacahRF} of the  Wilson rational functions and looking at the coefficients of, for $k=0,1,\dots {n+1}$, 
\begin{align}
   \tau_k(x) =\frac{(q^{-x},cq^{x+1};q)_k}{(aq^{-x}/c,aq^{x+1};q)_k}\,,
\end{align}
and using the following expression
\begin{align}
    &Z_{x+1,x}(c)\, \tau_k(x+1)+Z_{x,x}(c)\, \tau_k(x)   + Z_{x-1,x}(c)\, \tau_k(x-1)\nonumber\\
    =&
    \omega^+(k)\tau_{k+1}(x)
    +\omega^0(k)\tau_{k}(x)
    +\omega^-(k)\tau_{k-1}(x)\,.
\end{align}
We do not provide explicit expressions for $\omega^\pm(k),\omega^0(k)$ here, since they are not useful for the rest of this paper.



\paragraph{$R_{II}$ recurrence relation.}
 
Changing the normalization of the Wilson rational functions as follows
\begin{align}
 P_n(x)=\left(\frac{c }{a}\right)^n q^{-\binom{n}{2}}\ (aq^{x+1},aq^{-x}/c;q )_n\ W_n(x)\,, \label{RII}
\end{align}
it is easy to show, using the explicit expression of the ${}_{10}\phi_9$, that $P_n(x)$ is a polynomial of degree $n$ in terms of $\lambda(q^x;c)$.
As demonstrated in \cite{GM}, the polynomials $P_n(x)$ satisfies a $R_{II}$ recurrence relation. This relation can be computed straightforwardly from \eqref{eq:GM} remarking that
\begin{align}
 \frac{(aq^{x+1},aq^{-x}/c;q )_{n+1}}{(aq^{x+1},aq^{-x}/c; )_{n}}=(1-aq^{x+n+1})(1-aq^{n-x}/c )=\frac{aq^n}{c }\Big(\lambda(q^x;c) - \lambda( q^{-n-1}/a;c) \Big)\,.
\end{align}
In the notation used here, it reads
\begin{align}
 &Z_{n+1,n}(b)\, P_{n+1}(x) +\big( Z_{n,n}(b)\, \lambda(q^x;c)+X_{n,n}(b,c) \big) P_{n}(x)\nonumber\\
 +&Z_{n-1,n}(b)\, 
 \Big(\lambda(q^x;c) - \lambda(q^{-n}/a;c)\Big)
  \Big(\lambda(q^x;c) - \lambda(bq^{n}/a;c) \Big)P_{n-1}(x)  =0 \,.
\end{align}

\paragraph{Second type of GEVP.}By defining suitably a new rational functions $\overline{W}_n(x)$ through a change of the normalization of the Wilson rational functions, one can prove that $\oW_n(x)$ also satisfies a GEVP.
Let us define
\begin{align}
\oW_n(x)=\mu_n(x;b,c) W_n(x)\,,\label{Wbar}
\end{align}
with
\begin{align}
   \mu_n(x;b,c)=\frac{(aq^{x+1},aq^{-x}/c,bq/a,bcq^2/a;q)_n}{(bq^{1-x}/a,bcq^{x+2}/a,a/c,aq;q)_n} \,.
\end{align}
The renormalized Wilson rational functions $\overline{W}_n(x)$ still enjoy nice symmetry relations, for $n$ and $x$ non-negative integers,
\begin{align}
\oW_{x}(n)=\oW_n(x)\Big|_{b\leftrightarrow c}\,,
\end{align}
and very simple initial values $\oW_0(x)=\oW_n(0)=1$.

Using properties of the $q$-Pochhammer symbols, it is straightforward to show that $\overline{W}_n(x)$ satisfy a relation similar to \eqref{eq:GM}. Indeed, one finds 
\begin{align}\label{eq:GM2}
 \oA_n(x) \oW_{n+1}(x)   -\Big(\oA_n(x)+\oB_n(x)+\lambda(q^x;c)\bb_n(b,c)\Big)\oW_{n}(x)+ \oB_n(x) \oW_{n-1}(x)=0\,,
\end{align}
where 
\begin{align}
\bb_n(b,c)= \frac{q^{n-1}bc (q-aef)(q-ade)(q-adf)(1-cq/a)}{(c-aq^{n-1})(a-bcq^{2+n})}
\end{align}
\begin{subequations} \label{eq:Z2}
 \begin{align}
& \oZ_{n+1,n}{(b,c )}=\frac{qb(c-aq^{n})(1-b q^{n+1})(d- q^{n+1} )(e- q^{n+1})(f- q^{n+1})}
 { (1-b q^{2n+2})(1-b q^{2n+1})(a-bcq^{2+n})}\,,\\
 &\oZ_{n-1,n}{(b,c )}=\frac{ (a-bcq^{n+1})(1-q^n)(1- d b  q^{n})(1- e b  q^{n})(1- f b  q^{n})}
 { b(1-b q^{2n+1})(1-b q^{2n})(c-aq^{n-1})}\,,
\end{align}
\end{subequations}
and
\begin{align}
&\oA_n(x)= \oZ_{n+1,n}{(b,c )}\underbrace{(aq^x-bq^{n+1})(q^{-n-x-1}/b-cq/a)}_{\displaystyle =\lambda(q^x;c) - 
\lambda(bq^{n+1}/a;c)} \,,\\
&\oB_n(x)= \oZ_{n-1,n}{(b,c )} \underbrace{(1-aq^{x+n})(cq^{-n+1}/a-q^{-x} )}_{\displaystyle =\lambda(q^x;c) - 
\lambda(q^{-n}/a;c) } \,,
\end{align}
As for $W_n(x)$, a GEVP for $\oW_n(x)$ can be deduced from this relation.
\begin{prop}
The renormalized Wilson rational functions  $\oW_n(x)$ satisfy the following GEVP, for $n$ and $x$ non-negative integers:
\begin{align}
 &\oX_{n+1,n}(b,c)\, \oW_{n+1}(x)  +\oX_{n,n}(b,c)\, \oW_{n}(x)+ \oX_{n-1,n}(b,c)\, \oW_{n-1}(x)\nonumber\\
=&\lambda(q^x;c)\Big( \oZ_{n+1,n}{(b,c )}\, \oW_{n+1}(x)+\oZ_{n,n}{(b,c)}\, \oW_{n}(x)    + \oZ_{n-1,n}{(b,c )}\, \oW_{n-1}(x)\Big)\,, \label{eq:GEVPW3}
\end{align}
where 
\begin{align}
&\oZ_{n,n}{(b,c )}=-\Big(\oZ_{n+1,n}{(b,c )}+\oZ_{n-1,n}{(b,c)}+\bb_n(b,c)\Big)\,,
\end{align}
and
\begin{subequations}
    \begin{align}
& \oX_{n+1,n}(b,c)=\oZ_{n+1,n}{(b,c )}\, \lambda(bq^{n+1}/a;c) 
\,,\\
& \oX_{n-1,n}(b,c)=\oZ_{n-1,n}{(b ,c)}\,\lambda(q^{-n}/a;c)\,,\\
& \oX_{n,n}{(b,c )}= -\big(\oX_{n+1,n}(b,c)+\oX_{n-1,n}(b,c)\big)\,.
\end{align}
\end{subequations}
They also verify, for $n$ and $x$ non-negative integers:
\begin{align}
 &\oX_{x+1,x}(c,b)\, \oW_{n+1}(x)  +\oX_{x,x}(c,b)\, \oW_{n}(x)+ \oX_{x-1,x}(c,b)\, \oW_{n-1}(x)\nonumber\\
=&\lambda(q^n;b)\Big( \oZ_{x+1,x}{(c,b )}\, \oW_{n+1}(x)+\oZ_{x,x}{(c,b)}\, \oW_{n}(x)    + \oZ_{x-1,x}{(c,b )}\, \oW_{n-1}(x)\Big)\,. \label{eq:GEVPW4}
\end{align}
\end{prop}

For $n$ and $x$ non-negative integers, there exists a relation between the RHS of the recurrence and of the difference GEVP for $\oW_n(x)$ which is given by
 \begin{align}\label{eq:ZZs2}
&  \oZ_{n+1,n}(b,c)\, \oW_{n+1}(x)+\oZ_{n,n}(b,c)\, \oW_{n}(x)    + \oZ_{n-1,n}(b,c)\, \oW_{n-1}(x)\\
=&
\oZ_{x+1,x}(c,b)\,\oW_{n}(x+1)+\oZ_{x,x}(c,b)\, \oW_{n}(x)    + \oZ_{x-1,x}(c,b)\, \oW_{n}(x-1)\;.\nonumber
\end{align}
This relation is equivalent to \eqref{eq:ZZs}.

\section{Wilson rational algebra} \label{section:3}

Let us consider the tridiagonal matrices $X$ and $Z$ with the entries $\widetilde X_{i,j}$ (with $\rho=-(1+qc)$ in \eqref{eq:Xtilde}) and $Z_{i,j}$, respectively.
The $q$-commutator, the anticommutator and the $q$-number are defined as follows:
\begin{align}
[X,Z]_q=\frac{1}{1-q}(XZ-qZX)\,,\qquad \{X,Z\}=XZ+ZX\,,\qquad [x]=\frac{1-q^x}{1-q}\,.
\end{align}
For later convenience, let us introduce the symmetric function w.r.t. $e$, $d$ and $f$ given by
\begin{align}
\sigma(x)=1+e+d+f+\frac{1}{x}\left(1+\frac{1}{e}+ \frac{1}{d}+\frac{1}{f}\right)  \,.
\end{align}
We can show by direct computation that the matrices $X$ and $Z$ satisfy 
the following relations (we recall that the parameters are related by $bcdef=1$):
\begin{subequations}\label{eq:WilsonXZ}
\begin{align}
&[Z,[X,Z]_q]_q =\fc_1\{X,Z\}+\fc_2 Z^2+ \fc_3X +\fc_4 Z\,,\\
&[X,[Z,X]_q]_q =\fc_1 X^2+\fc_{2}\{X,Z\}+ \fc_{4}X+\fc_5 Z^3 +\fc_{6} Z^2 +\fc_{7} Z \,,
\end{align}
\end{subequations}
where the coefficients are given by
\begin{subequations}
\begin{align}
 \fc_1&=-\left(\frac{a}{bc}+\frac{q}{a}\right)\,,\quad \fc_2=  -q\sigma(b)\,,\quad
\fc_3=\frac{(1+q)^2}{bc}-\fc_1^2\,,\\
\fc_4&=
\frac{q(1+q)}{b}\sigma(c)-\fc_1\fc_2\,,\quad 
\fc_5 = (q+1/q)(q+1)^2c\,,\\
 \fc_{6}&=\frac{(1+q)[3]}{b}\big((1-b)(1-c)-bc\sigma(bc) -bc(1+1/q)\fc_1 \big)  
  -\frac{c(1-q^2)^2}{q} \fc_1 
\,,\\
 \fc_{7}&= -(1+q)^2c\fc_1^2
-\frac{[4]}{b}( (1-b)(1-c)-bc\sigma(bc) -bc\fc_1(1+1/q))\fc_1-q^2\sigma(-bq)\sigma(-b/q) \\
&
+(1+q)^2q(1+c/b)\sigma(bc)
-(1/(bc)+1)q(1+q)^2(1+c/b)
-(1-q^2)^2/b\,.\nonumber
\end{align}
\end{subequations}
The previous matrices $X$ and $Z$ encode algebraically the recurrence GEVP. 

Let us define the tridiagonal matrices $\oX$ and $\oZ$ with the entries $\oX_{i,j}$ and $\oZ_{i,j}$, respectively.
These matrices satisfy exactly the relations \eqref{eq:WilsonXZ}. This statement can be rephrased as $(X,Z)$ and $(\oX,\oZ)$ provide two different representation of the algebra defined by relations \eqref{eq:WilsonXZ}.

To take in account of the difference GEVP, one considers their eigenvalues $\lambda(q^n;b)$.
Let also define the diagonal matrix $V$ with non-vanishing entries $V_{n,n}=\lambda(q^n;b)$ and look for the relations between $V$, $X$ and $Z$. By straightforward computations, one shows that
\begin{align}\label{eq:WilsonVZ}
\frac{a}{bc}[ V,Z]_q +\frac{q}{a} [Z,V]_q
= \bd_1 V+\bd_2 X  +\bd_3 \un\,,
\end{align}
where $\un$ is the identity matrix,
and 
\begin{subequations}\label{eq:WilsonXV}
    \begin{align}
&[V,[X,V]_q]_q= \be_1 V^3 +\be_2 V^2+\be_3 X+\be_4 V+\be_5 \un\,,\\
&[X,[V,X]_q]_q= \be_6 VXV+\be_7\{V,X\}+\be_8 X+\be_{9} V+\be_{10}\un\,.
\end{align}
\end{subequations}
The coefficients in the previous relations are given by 
\begin{subequations}
\begin{align}
&\bd_1=\fc_3  \,, \quad \bd_2=(1+q)/c \,,\quad \bd_3=q\sigma(c)\fc_1+q\bd_2 \sigma(b)
\end{align}
\end{subequations}
and 
\begin{subequations}
\begin{align}
&\be_1=-(1+q)/b\,,\quad \be_2=\fc_2  \,,\quad \be_3=b(1+q)^2\,,\\
&\be_4=(1+q)^3 -\fc_1q\big( (b-1)(c-1)-bc\sigma(bc)\big)\,,\quad \be_5=bc(1+q)\bd_3\,, \\
&\be_6=\frac{[3]}{q}\be_1\,, \quad \be_7=\fc_2\,,\quad 
\be_8=\be_4 +\frac{(1+q)^3(1-q)^2}{q}
 \,,\quad \be_9=\fc_7+\frac{c(1-q^2)^2}{q} \fc_3
\,,\\
&\be_{10}=\frac{q^2}{b}\Big( (1+q)\sigma(c)  +\fc_1b\sigma(b)  \Big)\Big(bc\sigma(bc)-(1-b)(1-b)+bc\fc_1(1+1/q)\Big) +\frac{(1-q^2)^2c}{q}\bd_3\,.
\end{align}
\end{subequations}
The algebra generated by $X$, $Z$ and $V$
subject to the relations \eqref{eq:WilsonXZ}, \eqref{eq:WilsonVZ} and \eqref{eq:WilsonXV} will be called the Wilson rational algebra.

These  relations resemble the corresponding relations for the Askey-Wilson (or $q$-Racah) algebra. There are however additional terms in their rhs meaning that the above algebra has a more complicated structure. In particular, in contrast to the $q$-Racah algebra, there are no explicit expressions for the spectra of the operators $X$ and $Z$. In this respect the rational Wilson algebra seems to have a structure that is closer  to that of the $q$-Heun-Racah algebra, introduced in \cite{HeunAW,HeunRacah} and to the one appearing in the study of the Clebsch-Gordan problem of $sl(3)$ \cite{crampeE6,Missingsu3}. 

\section{Limits to $q$-hypergeometric ${}_4\phi_3$ \label{sec:4phi3} } \label{section:4}

In this section, limiting cases to ${}_4\phi_3$ functions presented in \cite{GM} are recalled and, for each case, the GEVP are presented.
Let us mention \cite{Bult} where these limits are classified and studied in detail. 

\subsection{$q$-Racah polynomial}

Let consider the limit $a\to \infty$. In this limit, $W_n(x)$ becomes
\begin{align}
W_n(x) \rightarrow R_n(x)= {}_{4}\phi_3 \left({{ q^{-n},\; b q^{n+1},\; q^{-x},\; c  q^{x+1}  }\atop
{  q/d,\; q/e,\; q/f }}\;\Bigg\vert \; q;q\right)\,,
\end{align}
with the constraint
\begin{align}
b c  def=1\,.
\end{align}
It is the usual definition of the $q$-Racah polynomials \cite{Koekoek}, when $f=q^{N+1}$ and $N$ a non-negative integers.
It is easy to see that the coefficients $X_{n+\epsilon,n}$ ($\epsilon=0,+1,-1$) scale as $\frac{1}{a^2}$, $Z_{n+\epsilon,n}$ ($\epsilon=+1,-1$) scale as $\frac{1}{a}$, ($\epsilon=+1,-1$) and $Z_{n,n}$ scales as $\frac{1}{a^2}$. Therefore, to leading order in the limit $a\to 0$, the GEVP \eqref{eq:GEVPW1} becomes the following EVP:
\begin{subequations}
\begin{align}
 &A_n R_{n+1}(x)  -(A_n+C_n)R_{n}(x)+ C_n R_{n-1}(x)=-\lambda(q^x;c)R_{n}(x) \,,  \label{eq:EVPR}
\end{align}
where
\begin{align}
&A_n=\frac{(1-b q^{n+1})(1- q^{n+1}/d )(1- q^{n+1}/e)(1- q^{n+1}/f)}
 {(1-b q^{2n+2})(1-b q^{2n+1})}\,,\\
&C_n=\frac{c  q(1-q^n)(1- d b  q^{n})(1- e b  q^{n})(1- f b  q^{n})}
 {(1-b q^{2n+1})(1-b q^{2n})}\,.
\end{align}
\end{subequations}
The relation above is the well-known recurrence relation of the $q$-Racah polynomials \cite{Koekoek}.
The difference equation satisfied by the $q$-Racah polynomial $R_n(x)$ can be obtained similarly using \eqref{eq:GEVPW2}.

\subsection{First $q$-Racah rational function}

Let us consider the limit $c \to \infty$ and $ f \to 0$ with  $fc\to \frac{1}{b de }$.
In this limit the Wilson rational functions become: 
\begin{align}
W_n(x)\to \sR^{(1)}_n(x)={}_{8}\phi_7 \left({{ a,\;q\sqrt{a},\;-q\sqrt{a},\; q^{-n},\; b q^{n+1},\; q^{-x},\; ad  ,\;ae }\atop
{\sqrt{a},\;- \sqrt{a},\;  aq^{-n}/b ,\;aq^{n+1},\;   aq^{x+1},\; q/d,\; q/e }}\;\Bigg\vert \; q;\frac{q^{x+1}}{b de}\right)\,.
\end{align}
The Watson's transformation formula (see relation (III.18) in \cite{GR}) allows to transform a ${}_8\phi_7$ to a ${}_4\phi_3$ as follows:
\begin{align}
&{}_{8}\phi_7 \left({{a,\;q\sqrt{a},\;-q\sqrt{a},\; b,\; c,\;  d  ,\;e,\;q^{-n} }\atop
{\sqrt{a},\;- \sqrt{a},\;   aq/b,\;  aq/c,\; aq/d,\; aq/e,\;aq^{n+1} }}\;\Bigg\vert \; q;\frac{a^2q^{n+2}}{bcde}\right)\nonumber\\
=&\frac{(aq,aq/(de);q)_n}{(aq/d,aq/e;q)_n}\ {}_{4}\phi_3 \left({{aq/(bc),\; d,\;  e, \;q^{-n} }\atop
{aq/b,\;  aq/c,\; deq^{-n}/a }}\;\Bigg\vert \; q;q\right)\,.\label{eq:Watson}
\end{align}
Using this formula for $a\to a\,, b\to ad\,,\ c\to ae\,,\ d\to bq^{n+1}\,,\ e \to q^{-x}\,,\ q^{n}\to q^n$, one gets, for $n,x$ non-negative integers:
\begin{align}\label{eq:qR2bis}
\sR^{(1)}_n(x)=\frac{(aq,aq^{x-n}/b;q )_n}{(aq^{x+1},aq^{-n}/b;q )_n} {}_{4}\phi_3 \left({{q^{-n},\; b q^{n+1},\; q^{-x},\; q/(ade) }\atop
{q/d,\;q/e,\;   b q^{1-x}}/a}\;\Bigg\vert \; q;q\right)\,.
\end{align}
In this limit, $\sR^{(1)}$satisfies a GEVP but the associated Wilson rational algebra simplifies drastically. Indeed, let us denote by $\cX$, $\cZ$ and $\cV$ the leading terms of the matrices $X$, $Z$ and $V$, respectively. To simplify the presentation, we renormalize the matrix $\cZ$ by a factor $q/(bde)$. These matrices satisfy the following relations:
\begin{subequations}\label{eq:meta}
\begin{align}
 &[\cX,\cZ]_q=  \ff_1\cZ+\ff_2\cX \,,\\
  &[\cZ,\cV]_q=\ff_2 \cV+ \ff_4\cX+\ff_5\un\,,\\
 &[\cV,\cX]_q= \ff_1 \cV+\ff_6 \cZ+ \ff_7 \un\,,
\end{align}
\end{subequations}
where
\begin{subequations}\label{eq:f}
\begin{align}
& \ff_1=-de\,,\quad \ff_2= -q/a \,,\quad 
\ff_3= 1+bq-a(1+q)\,,\\
&\ff_4=(1+q)a\,,\quad \ff_5= ade(1+q)-q(bed+e+d+1) \,,\\ &\ff_6=\frac{b}{a} (1+q)\,,\quad
\ff_7=\frac{qb}{a^2}(1+q)-\frac{q}{a}(bed+eb+bd+1) \;.
\end{align}
\end{subequations}
The algebra generated by $\cX$, $\cZ$ and $\cV$ subject to the relations \eqref{eq:meta} stands to be the meta $q$-Racah algebra that will be denoted by $m\fR_q$. We expect that the meta $q$-Racah algebra generalizes the meta $q$-Hahn algebra and  meta Hahn algebra studied in \cite{tsujimoto2024meta}-\cite{bernard2024meta} and completes the set of this type of algebra..

Performing the same limit on the renormalized Wilson rational functions $\oW_n(x)$, one obtains: 
\begin{align}\label{eq:qR2ter}
\oW_n(x) \to \osR^{(1)}_n(x)= {}_{4}\phi_3 \left({{q^{-n},\; b q^{n+1},\; q^{-x},\; q/(ade) }\atop
{q/d,\;q/e,\;   b q^{1-x}}/a}\;\Bigg\vert \; q;q\right)\,,
\end{align}
and denote by $\ocX$, $\ocZ$ and $\ocV$ the leading terms of the matrices $\oX$, $\oZ$ and $V$, respectively. These matrices satisfy the following relations:
\begin{subequations}\label{eq:meta2}
\begin{align}
 &[\ocZ,\ocX]_q=\ff_1\ocZ+\ff_2\ocX\,,\\
 &[\ocX,\ocV]_q=\ff_2 \ocV+\ff_4 \ocX+\ff_5 \un\,,\\
 &[\ocV,\ocZ]_q=\ff_1 \ocV+\ff_6 \ocZ+ \ff_7\un\,,
\end{align}
\end{subequations}
where the coefficients are given by \eqref{eq:f}.
They are the same relations than relations \eqref{eq:meta} but with the generators exchanged on each l.h.s.

\subsection{Other $q$-Racah rational functions}

Let us consider the limit $c  \to \infty$ and $b \to 0$ with $ b c  \to \frac{1}{def}$. In this limit, the Wilson rational functions $W_n(x)$ become (after using formula \eqref{eq:Watson}, for $a\to a\,, b\to ad\,,\ c\to af\,,\ d\to q^{-x}\,,\ e \to ae\,,\ q^{n}\to q^n$):
\begin{align}\label{eq:qRacahR12}
\sR^{(2)}_n(x)=\frac{(aq,q^{x+1}/e;q)_n}{(aq^{x+1},q/e;q)_n}\ {}_{4}\phi_3 \left({{ q^{-n},\; q^{-x},\;  ae  ,\;q/(adf) }\atop
{  q/d,\; q/f,\;eq^{-x-n} }}\;\Bigg\vert \; q;q\right)\,.
\end{align}

The previous limit on $\oW_n(x)$ goes to 
\begin{align}\label{eq:qRacahR13}
\oW_n(x)\to \osR^{(2)}_n(x)=\frac{(q^2/(adef),q^{x+1}/e;q)_n}{(q^{x+2}/(aedf),q/e;q)_n}\ {}_{4}\phi_3 \left({{ q^{-n},\; q^{-x},\;  ae  ,\;q/(adf) }\atop
{  q/d,\; q/f,\;eq^{-x-n} }}\;\Bigg\vert \; q;q\right)\,.
\end{align}
For this two cases, the Wilson rational algebra does not simplify. It is easy to compute the coefficients in this limit. The results are not particularly enlightening and we do not record them here.

There exists another limit providing another $q$-Racah rational functions.
Indeed, considering the limit $e \to \infty$ and $f\to 0$ with $ ef \to \frac{1}{b c  d}$ and using Watson's transformation formula \eqref{eq:Watson}, for $a\to a\,, b\to q^{-x}\,,\ c\to cq^{x+1}\,,\ d\to bq^{n+1}\,,\ e \to ad\,,\ q^{n}\to q^n$), the Wilson rational functions become:
\begin{align}
W_n(x)\to \sR^{(3)}_n(x)=\frac{(aq,q^{-n}/(bd);q)_n}{(aq^{-n}/b,q/d;q)_n} {}_{4}\phi_3 \left({{  q^{-n},\; b q^{n+1},\; a/c,\; ad  }\atop
{aq^{x+1},\;aq^{-x}/c,\; bdq}}\;\Bigg\vert \; q;q\right)\,.
\end{align}
The Wilson rational algebra associated to this functions has the same structure as the one for the Wilson rational function but numerous coefficients vanish: $\fc_2,\fc_4,\fc_5,\fc_{10},\bd_2,\be_1,\be_7 \to 0$.

\section{Concluding remarks} \label{section:5}

In the framework of a program aimed at extending the Askey scheme to BRF and at providing through the so-called meta algebras a representation theoretic synthesis of the bispectral properties of these functions, we set out to identify the meta $q$-Racah algebra whose definition was still missing. This goal was achieved by analysing the features of the Wilson rational functions from that same perspective and taking limits. This had the remarkable outcome of leading to the definition of the rational Wilson algebra from which the meta $q$Racah algebra was obtained. 

The results presented here open various research avenues many of which having to do with the second route mentioned in the introduction where the starting point are the algebras instead of the functions. First it paves the way for the completion of the unified algebraic treatment of the hypergeometric OPs and BRFs through the construction that can now be done of the representations of the $q$-meta Racah algebras and its $q=1$ version that were identified. Second, from the knowledge of the various examples of meta algebras that have been defined and used, it would be of significant interest to develop the general theory of this class of algebras. Finally, it would be a remarkable accomplishment to show that the rational Wilson algebra encodes the bispectral properties of the eponymous functions by showing that the recurrence coefficients  of the GEVP they satisfy can be recovered from the representations of this algebra.

\paragraph{Acknowledgements:}
N.~Cramp\'e is partially supported by the international research project AAPT of the CNRS. The research of S.~Tsujimoto is supported by JSPS KAKENHI (Grant Number 19H01792). L.~Vinet is funded in part by a Discovery Grant from the Natural Sciences and Engineering Research Council (NSERC) of Canada.

\end{document}